\documentclass[12pt]{article}
\usepackage{cite}

\newcommand{\abs}[1]{\left | #1 \right |} 
\newcommand{\beq}{\begin{equation}}
\newcommand{\eeq}{\end{equation}}

\newcommand{\lam}{\lambda}
\newcommand{\vev}[1]{\left < #1 \right >}
\newcommand{\ra}{\rightarrow}

\newcommand{\nc}{\newcommand}

\begin{document}

\begin{titlepage}

\begin{center}

%\vspace{2cm}

{\hbox to\hsize {\hfill PUPT-1868, MIT-CTP-2875}}
{\hbox to\hsize{\hfill {hep-ph/9906296}}
\bigskip

%\vspace{2cm}

\bigskip

\bigskip

\bigskip

{\Large \bf  Heavy Thresholds, Slepton Masses and
the $\mu$ Term in  
Anomaly Mediated Supersymmetry Breaking}

\bigskip

\bigskip

{\bf Emanuel Katz}$^{\bf a}$,
 {\bf Yael Shadmi}$^{\bf b, c,}$\footnote{On leave of absence
from the Weizmann Institute of Science, Rehovot, Israel.}, 
and  {\bf Yuri~Shirman}$^{\bf b, d}$ \\

\bigskip

\bigskip

$^{\bf a}${\small \it Center for Theoretical Physics\\Department of
Physics\\MIT, Cambridge, MA 02139 \\
{\rm email}: amikatz@mit.edu }
}
\smallskip

 \bigskip

$^{\bf b}${ \small \it  Department of Physics\\ Princeton University\\ 
Princeton, NJ 08544, USA\\

$^{\bf c}${\rm email}: yshadmi@feynman.princeton.edu\\ 

$^{\bf d}${\rm email}: yuri@feynman.princeton.edu \\
}

\vspace{1cm}

{\bf Abstract}

\end{center}
\noindent
The effects of heavy mass thresholds on anomaly-mediated
soft supersymmetry breaking terms are discussed.
While heavy thresholds completely decouple
to lowest order in the supersymmetry breaking,
it is argued that they do affect the breaking
terms at higher orders.
The relevant contributions typically occur at 
lower order in the loop expansion compared to
purely anomaly mediated contributions.
The non decoupling contributions may be used to
render models in which the only source of
supersymmetry breaking is anomaly mediation
viable, by generating positive contributions
to the sleptons' masses squared.
They can also be used to generate acceptable $\mu$-
and $B$-terms.

\end{titlepage}

\renewcommand{\thepage}{\arabic{page}}
\setcounter{page}{1}

\section{Introduction}

Recently, it was pointed out that in the presence of
any supersymmetry-breaking hidden sector, soft supersymmetry-breaking terms
are generated for the observable fields through the
super-Weyl anomaly~\cite{rs,rat}. Because of their origin, these
supersymmetry (SUSY) breaking terms are directly 
related to the scaling dimension
of the relevant operator.

The soft terms are most readily obtained by working in
a supergravity formulation  that uses the superconformal, 
or super-Weyl compensator, $\Phi$, a non-dynamical chiral superfield
which allows one to write a manifestly super-Weyl invariant
action at the classical level~\cite{kl}.
Essentially the role of $\Phi$ is to compensate for the non-trivial
super-Weyl transformations of different action terms,
while any $\Phi$ vev breaks the super-Weyl invariance.
In particular, one may choose the lowest component $\Phi$
vev to be 1,  leading to a trivial modification of the action.
However, for the discussion of supersymmetry breaking, we will
focus on the auxiliary component of $\Phi$. As was argued in~\cite{rs},
in the presence of supersymmetry breaking, this auxiliary component 
acquires a non-zero vev, which we denote by $F$, so that 
$\Phi \equiv 1+ F \theta^2$.

To recover the observable sector Lagrangian coupled to
canonical supergravity, one then rotates $\Phi$ away
through a super-Weyl rescaling. For a scale-invariant
observable sector Lagrangian, $\Phi$ disappears altogether.
However, if any explicit mass scale appears in the Lagrangian,
it gets rescaled by $\Phi$ as $M\ra M \Phi$, so that in the presence of
supersymmetry breaking, various tree-level supersymmetry splitting
masses appear.

Even if there are no explicit mass scales in the classical
theory, some mass scale is generated quantum mechanically.
In particular, if we consider the renormalized Lagrangian,
the wave function renormalizations of the fields contain some
cutoff dependence. The cutoff scale now appears multiplied by $\Phi$.
Expanding the wave function renormalization in $F$ (which enters
through $\Phi$),
one finds a soft supersymmetry breaking mass for the 
relevant field.\footnote{The extraction of these soft masses is
completely analogous to the method of calculating
soft masses in models of gauge-mediated supersymmetry breaking
through wave function renormalizations of~\cite{gr,gr1}.}

Let us concentrate for now on the scalar masses squared.
Since these masses squared enter through~\cite{rs}
\beq
\label{mz}
Z(\mu) = Z_0(\mu)\,\left(1 -{1\over 2} \gamma_i(\mu)\, (F\theta^2+c.c.) 
+ {1\over 4} 
\stackrel{\cdot}{\gamma}_i(\mu)\, 
\abs{F}^2\theta^2{\bar\theta}^2\right)\ ,
\eeq
they have two immediately apparent features. First, the soft mass has
no ${\cal O}(F^3)$ or higher contributions. Second, while the soft
masses can be obtained through~(\ref{mz}) at the UV cutoff of the
theory and then evolved down to any lower scale, it is also possible
to directly evaluate them at the low scale, again through~(\ref{mz}).
The reason is of course that the wave function renormalization
already takes running effects into account.
Thus, at any scale, the anomaly-mediated (AM) masses
appear to be determined solely by the theory at that
scale, specifically by the $\beta$-function and anomalous dimensions
at that scale, with no memory of the theory at higher scales.
This picture has been derived to leading order in the SUSY
breaking parameter in a theory in which the only source of
SUSY breaking is anomaly mediation. 

If there are any additional contributions to the soft masses
at high scales (such as the compactification scale),
from, say,  different gravitational sources, these contributions
affect the running of the masses to lower scales.

Even if the only source of supersymmetry breaking
in the theory is anomaly mediation, which generates 
soft masses near the UV cutoff of the theory which are
given by~(\ref{mz}), one may wonder about what happens when the
theory also contains some heavy thresholds. 
The mass of the heavy fields, which we denote by $M$, is rescaled
as $M\ra M \Phi$. Therefore the heavy fields obtain 
tree-level supersymmetry splittings proportional to $F$.
If the heavy fields interact with some of the light fields,
they can then contribute to the light fields' soft masses
at the loop level.
Still, these contributions largely decouple~\cite{rs,rp}.
More specifically, the contribution of any heavy field decouples
at order~$F^2$.
This should be apparent from our discussion above -- since the
AM soft masses can be read of the wave-function renormalizations,
the effects of heavy thresholds should somehow be accounted for
already. In section~2, we will show in detail how this decoupling
occurs. 
%We will also point out two effects leading to
%non-decoupling of the heavy thresholds.
However, we will argue that the heavy thresholds do not
decouple completely. There are in fact two sources of non-decoupling.

First, the ${\mathcal O} (F^2)$ effects of the heavy threshold 
do not decouple if there is a light field associated with this 
threshold \cite{rp}. More specifically, this happens if the masses of the 
heavy fields are determined by some modulus, and this modulus is
stabilized primarily by small supersymmetry breaking effects.

Second, if there are fields of mass $M$ and
supersymmetry splittings proportional to $F$ that interact with the
light fields, the loop-level soft masses that they induce
contain contributions that are higher order in $F$. For example,
scalar masses squared obtain contributions of order $F^4/M^2$.
Furthermore, in many cases these contributions start at one-loop,
whereas the $F^2$ contributions start at two-loops. Thus, if $M$ is not
much bigger than $F$, these contributions may be important.

Another interesting issue to address is the decoupling of D-terms.
As we shall see in section~3, in anomaly mediation D-terms decouple 
to leading order in the SUSY breaking, unlike in the usual case. 
Again, however, $F^4/M^2$ contributions to D-terms do not decouple.

The above observations may be used to address the main phenomenological
problem of the minimal anomaly-mediation scenario~\cite{rs}. 
In this scenario, the soft breaking terms of the supersymmetric standard
model (SM) are generated purely by anomaly mediation, and are
therefore given by~(\ref{mz}).
The slepton masses squared are proportional to the $SU(2)$,
and $U(1)$ $\beta$ functions, and are therefore negative.
Ref.~\cite{rs} invoked additional gravitational contributions
to overcome this problem. Ref.~\cite{rp} proposed a solution
in which the soft spectrum is modified by the presence
of a light (order $F$) modulus, which is massless in the
supersymmetric limit. Another solution involves
additional Higgs doublets that generate large Yukawa couplings for
the sleptons~\cite{luty}.
Here we will instead introduce new heavy thresholds, within 
one or two orders of the supersymmetry breaking scale $F$, 
and utilize the resulting $F^4$ soft masses to generate
positive slepton masses squared. Our model, which has as its basis
a simple $U(1)$ theory, which we describe in Section~3,
illustrates the
decoupling of order $F^2$ contributions and the non-decoupling
of $F^4$ terms. It also clarifies the issue of D-term decoupling
in AM scenarios, as some of the contributions we will discuss
can be understood as arising from the $U(1)$ D-term.

As we said above, the crucial contributions in our model
are of the form $F^4/M^2$. In Section~4, we will show
how, in a modification of our model, the relevant scale $M$  
is generated dynamically from $F$. 
In fact, as we will see, it is quite easy to generate
scales that are close to the SUSY breaking scale using
anomaly mediation. This fact could be useful for model building
purposes.
In Section~5,
we put these pieces together and construct a modification
of the SM in which sleptons obtain positive masses squared.

In Section~6, we again use the same ingredients,
that is, non-decoupling ${\cal O}(F^4/M^2)$ contributions
and a scale $M$ that is generated dynamically from $F$,
to naturally obtain a $\mu$-term and a $B$-term of the correct
size.

We close this Introduction with one comment. The mechanism
of anomaly-mediated supersymmetry breaking exists in any
theory with some supersymmetry breaking sector. In particular,
it exists in any 4-dimensional theory. However, in a 4-dimensional
theory, higher order, $M_p$-suppressed operators that couple
the observable and hidden sector, typically generate tree-level
soft masses for the observable fields. These are the well known
``hidden sector'' contributions. These contributions are larger
than the anomaly mediated contributions, which are loop-suppressed.
In a 4-dimensional theory, it is hard to rule out the existence
of the tree-level hidden sector contributions. However, in~\cite{rs},
it was shown that in $d>4$ dimensions, such direct couplings 
of the observable and hidden sector can plausibly be absent.
Thus, while our discussion of decoupling is independent
of the dimensionality of the full theory, when we move on
to model building we envision a situation in which
the soft terms are purely anomaly mediated, as in
sequestered sector models \cite{rs}. We implicitly
assume then that the full theory has more than 4 dimensions,
and the appropriate cutoff scale, at which the anomaly-mediated
soft masses of the observable sector are generated, is
the compactification scale.

\section{The effect of mass thresholds on soft masses of light fields}

We would like to study the impact of supersymmetric mass thresholds 
on anomaly mediated SUSY breaking terms.  
We assume that the heavy thresholds lie above the
visible sector SUSY breaking scale, $F$, and below the UV cutoff scale,
whether it is the Planck scale or the compactification scale.
The decoupling of these thresholds can then
be addressed in terms of an effective 4-dimensional field theory. 
We also assume that 
there are no sources of SUSY breaking other than anomaly mediation.  
A discussion of decoupling effects at leading
order in the SUSY breaking recently also appeared in
Ref.~\cite{rp}.

In the absence of heavy thresholds, 
slepton and squark soft masses follow directly 
from the wave function renormalization~\cite{rs}:
\begin{eqnarray}
&& \int d^4 \theta Z\left(\frac{\mu}{\Lambda 
(\phi^\dagger \phi)^{1/2}}\right) Q^\dagger 
Q \\ \nonumber
&=& \int d^4 \theta \,
Z\left(\frac{\mu}{\Lambda}\right)\,\left( 1 - \frac{1}{2} \gamma (F 
\theta^2 +  c.c) +  \frac{1}{4} (\partial_t \gamma) \abs{F}^2 \theta^2 
\bar{\theta}^2 \right) Q^\dagger Q\ .
\end{eqnarray}
Here $\Phi = 1  + F \theta^2$ is the super-Weyl compensator, $\Lambda$ is 
the UV cutoff, $t=\ln \mu$, and the anomalous dimension 
$\gamma=\partial\ln Z/ \partial \ln\mu$. 
Rescaling the fields 
\beq
Q \rightarrow Z^{-1/2}\exp{(\frac{1}{2} \gamma F \theta^2)}\, Q\ , 
\eeq
to obtain canonical kinetic terms,
and expanding the remaining exponent 
yields the following soft mass for the light field $Q$:
\beq
m_s^2 = -\frac{1}{4} \abs{F}^2 (\partial_t \gamma) \ .  
\eeq
We see that the soft masses are determined by the anomalous
dimensions of the light fields.
Thus, 
the soft masses (as well as other soft parameters) 
are largely insensitive to the details of  
the high energy theory.

However, in the presence of heavy mass thresholds, there may be additional
contributions to the soft masses of the light fields.
The theory contains explicit mass parameters 
which determine the heavy threshold.  
Due to the scaling anomaly, all such  parameters will
be accompanied by the compensator field $\Phi$.  
The heavy fields then acquire tree-level supersymmetry-breaking
mass splittings. 
As these fields are integrated out, their supersymmetry-breaking 
splittings may in principle generate soft terms for the light
fields, through gauge or Yukawa interactions.
We would like to understand to what extent 
such thresholds
affect the masses of the light fields,
and in particular, whether their effects completely decouple.

To study these questions, it will prove convenient
to distinguish between two possibilities.
One is that the heavy threshold, or in other words,
the mass of the heavy fields associated with this threshold, 
appears as an explicit mass term in the Lagrangian.
The second is that the mass of the heavy field
is given by the expectation value of some dynamical field.
These two cases are different because vevs are determined
dynamically and may depend on non-supersymmetric effects.

Let us first consider the case that the masses of 
all heavy fields arise from 
explicit mass terms in the superpotential. 
As explained above, any mass term $M$ should be promoted
to a superfield-valued $X=M\Phi$.
We can then obtain the soft masses of the light fields
at a scale $\mu < M$ from the relevant wave function renormalization
$Z(\mu)$. 
Assume for simplicity that $Z$ depends on a single
coupling.  
Then, from the 1-loop relation 
\beq
\label{z}
\partial_t Z = \frac{\alpha}{\pi}c\ ,
\eeq
where $c$ is the appropriate quadratic Casimir,
we first solve~(\ref{z}) for $Z$ in the low energy theory,
and  then match boundary conditions at $M$.
Using  arguments of holomorphy and 
R-symmetry~\cite{gr,gr1}, we then make the substitutions 
$M^2 \rightarrow X^\dagger X$
and $\Lambda^2 \rightarrow \Lambda^\dagger \Lambda$.  
Note that both $X$ and $\Lambda$ are superfield valued
and contain the compensator $\Phi$.
This yields
\beq
Z\left(\alpha(\mu,X), \alpha(X)\right),
\eeq  
where, 
\begin{eqnarray}
\label{alphaH}
\alpha(X)^{-1} &=& \alpha(\Lambda)^{-1} 
+ \frac{b_H}{4\pi}\ln\frac{X^\dagger X}{\Lambda^\dagger
\Lambda} 
=
\alpha(\Lambda)^{-1} +
\frac{b_H}{4\pi}\ln\frac{M^2}{\Lambda^2} \ ,\\
\label{alphaL}
\alpha(\mu,X)^{-1} &=& 
\alpha(X)^{-1} + \frac{b_L}{4\pi}
\ln\frac{\mu^2}{X^\dagger X} \\ \nonumber
&=& 
\alpha(\Lambda)^{-1} + \frac{b_L}{4\pi}
\ln\frac{\mu^2}{\Lambda^2 \Phi^\dagger\Phi}
+ \frac{b_H-b_L}{4\pi}\ln\frac{M^2}{\Lambda^2} 
\ , 
\end{eqnarray}
where $b_H$ ($b_L$) is the one-loop beta function above
(below) the scale $M$.
Note that 
the $\Phi$ dependence cancels between 
$X$ and $\Lambda$ in the last term of~(\ref{alphaL}).  
Therefore, the only contribution to the soft-mass squared,
which is ${\cal O}(F^2)$, comes from the second term in
the second line of eq.~(\ref{alphaL}).
Since this contribution is not related to the threshold $M$,
the low-energy theory is 
insensitive to the 
mass splittings associated with the heavy threshold.

It is important to stress that while the previous
discussion can be generalized to all orders in the coupling
constants, it
only holds to leading order in supersymmetry breaking,
that is, ${\cal O}(F^2)$ corrections to the scalar-masses
squared.
We will return to the question of higher order corrections
in the end of this section.

As we just saw, the crucial point for the decoupling of the heavy fields
is that their mass scale $M$ is rescaled by $\Phi$ in
the same way that $\Lambda$ is, so that the $\Phi$ dependence
cancels in the ratio. This implies a specific relation between
the SUSY masses and the soft masses of the heavy fields.
We will refer to masses that are obtained by the rescaling
of some mass-scale by $\Phi$ as being ``aligned''.
For example, the mass of a  chiral field is aligned
if
\beq
\frac{\abs{m_{scalar}^2-m_{fermion}^2}}{m_{fermion}} = F  ~,
\eeq
and  ${\rm Str} m^2 = 0$ in a supermultiplet.

The decoupling of the heavy threshold which we just saw
may be understood as a cancellation
between the anomaly-mediated contribution and
the contribution of the heavy fields.
To see that explicitly, consider an example in which the
heavy fields are $N_f$ pairs of fundamentals and antifundamentals
of an $SU(N)$ gauge group, with supersymmetric mass $M$.
We also assume that the light fields transform as fundamentals
of the same $SU(N)$.
Consider then the soft masses of the light fields,
evaluated at a scale $\mu > M$, 
\beq
\label{am}
m^2(\mu)_{\rm anomaly-mediated} = 2 c_0
b_H \,{\alpha^2 \over 16\pi^2}\, F^2\ , 
\eeq
with $b_H$ and $c_0$ the one-loop beta function  
and one-loop anomalous dimension coefficients, respectively.
To evaluate these masses at low-scales, we should evolve
them down to the low-scale, but the only thing that evolves
in~(\ref{am}) is the coupling $\alpha$.
However, as we go below the threshold $M$, we
also need to integrate out the heavy fields,
and they contribute to the soft masses of the light
fields. Their contribution is the usual gauge-mediated
contribution,
\beq
\label{gm}
m^2(\mu)_{\rm gauge-mediated} = 2 c_0
 N_F\,{\alpha^2 \over 16\pi^2}\, F^2\ , 
\eeq
Adding the contributions~(\ref{am}) and~(\ref{gm}), 
we find, 
at $\mu < M$,
\beq
\label{tot}
m^2(\mu) = 
2 c_0\, b_L\,{\alpha^2 \over 16\pi^2}\, F^2\ , 
\eeq
since $b_L = b_H + N_f$. This is precisely the soft mass
we would obtain when calculating directly 
in the low-energy theory below the threshold $M$,
ignoring the heavy fields altogether.
We see that the contribution of the heavy fields
through anomaly mediation, which is proportional
to $N_f$ and arises through their contribution
to the beta function, exactly cancels their contribution
through gauge mediation. Our discussion can be easily
generalized to models with Yukawa couplings between light
and heavy fields. 
 
This cancellation may also be seen diagrammatically in the
Pauli-Villars regularization scheme.  The regulator fields  
always have aligned masses $\Lambda \Phi$, because an explicit mass
term must be put in for them.  Now, when we consider any diagram 
contributing to the soft masses, with heavy fields running in the 
loop, there will be an analogous diagram with the regulator 
fields running in the loop.  We can assume that the real fields have 
mass $M$, the regulators have mass $\Lambda$, and treat the splittings
$MF$ and $\Lambda F$ as insertions.  
Then, to order
$F^2$ (for which there is no dependence on the heavy mass), 
the heavy fields and
the regulator fields have 
exactly 
opposite contributions to the soft masses.  
The contributions are guaranteed to 
be of 
exactly the
same magnitude because 
the masses 
of both the heavy fields and the regulators
are aligned.

Now let us consider 
the possibility that
some SUSY masses originate from non-zero vevs.  
Take $X$ to be a field which acquires a vev in the supersymmetric
limit. Fields that couple to $X$
obtain a SUSY preserving mass $C \vev{X}$, and soft mass
splittings $C F_X$, with $C$ some constant
that depends on different 
couplings.\footnote{For simplicity we assume here
that the coupling is renormalizable, but it is easy to
repeat the following discussion for the general case.} 
Such heavy fields  decouple from the
effective theory of the light degrees of freedom if 
their masses are aligned, that is, if
$F_X = X F$.
In the limit of unbroken SUSY, we write  
$X = v +\delta  x$, where
$v$ is a background superfield whose lowest component is
just the vev of $X$, while $\delta x$ is the fluctuating part of
$X$ whose vev vanishes in the supersymmetric limit. 
Since $v$ is determined by a
combination of  mass parameters in the superpotential,
upon supersymmetry breaking, it acquires an F-term 
which is
automatically aligned with the scalar component, 
$v \ra v (1+F \theta^2)$. 
On the other hand, 
when supersymmetry is broken, $\delta x$ can obtain 
both a scalar vev and an auxiliary component vev,
that are determined by non-holomorphic effects and 
thus are not necessarily aligned.
Thus we need to
understand under which conditions the  vevs can become 
significantly non-aligned.
The Lagrangian for $\delta x$ has the form
\begin{eqnarray}
\label{xW}
&\int d^4 \theta ~
\left[(1- m_x^2 \theta^4) (\delta x)^\dagger (\delta x) +
\left (v(\Phi^\dagger - m_x^2 \theta^4) (\delta x) 
+ {\rm c.c.}\right ) \right ]+\\ \nonumber 
&\left( \int d^2 \theta 
\frac{1}{2}\, m\,(\Phi - a \theta^2) (\delta x)^2 + {\rm c.c.}\right)  ~,
\end{eqnarray}
where $m$ is the mass of $\delta x$ in the
supersymmetric limit, while $a$ and $m_x^2$ 
are soft mass parameters generated 
by anomaly mediation.
Note that while we can treat $m$ 
as a free parameter (in a given model, it 
is determined by the detailed form of $W(X)$), 
$a$ and $m_x$ are suppressed relative to $F$  by a 
one loop factor, $a\sim m_x \ll F$.
We see that the radiative soft parameters in
the Lagrangian will 
modify $X$ and $F_X$.
Solving 
for $F_{\delta x}$ 
and  extremizing the potential we find for the corrected
values of $X$ and $F_X$ at the extremum
\beq
\label{Xsol}
X =  v\frac{ m^2+ m a}{m^2 +m F + m_x^2 + ma}\ ,
~~~~~F_X= v\frac{m^2 F + m m_x^2}{m^2+mF + m_x^2 + ma} \ .
\eeq
It is clear that in the limit $m \ra \infty$ we recover
the results obtained for explicit mass parameters in the
Lagrangian as should have been expected.
On the other hand, in the limit $m \ra 0$ the extremum 
is found at $X=F_X=0$
independently of the value of $m_x^2$. This point may in
fact be the maximum of the potential if $m_x^2 <0$. In 
such a case there may exist a local minimum at non-zero $X$
if $m_x^2$ changes sign as a function of $v$
\cite{rp}. Moreover, a local minimum may exist at non-zero
$X$ for any sign of $m_x^2$ if cubic and higher order
superpotential terms for $\delta x$
are taken into account \cite{rp}. 
In any case, as we show and quantify momentarily, 
for small $m$ alignment is
lost in a vacuum with non-vanishing $X$.

Assuming that the eigenvalues of the $X$
mass matrix are positive, so that~(\ref{Xsol}) gives a
minimum of the potential, we can easily check whether
the corrected values of $X$ and $F_X$ are aligned,
\beq
\label{hierarchy}
\frac{F_X}{X} = F ( 1 -\frac{a}{m}+  
\frac{a^2}{m^2} + \frac{m_x^2}{m F} + \ldots) \ ,
\eeq
where the dots stand for higher order terms in $m_x^2$ and $a$.
We see that alignment is approximately preserved if
\beq
\label{aligncond}
a \sim \frac{y^2}{16 \pi^2} F \ll m \ , 
\eeq
where $y$ stands for the couplings of $X$. Note that for
small $m$, $m_x^2$ gives
a subleading contribution since it 
appears at two loops whereas $a$ appears at one loop.
Moreover, if  $X$ interacts sufficiently
weakly (so that $a$, which is one loop suppressed, 
is sufficiently small),
it is possible that the mass of $\delta x$ is significantly below 
the SUSY breaking scale, $m \ll F$, yet alignment holds to a good
approximation.
We therefore conclude that heavy fields whose masses are
generated by the vevs of dynamical fields decouple
so long as all mass parameters (including the masses of the fields which
obtain vevs) satisfy the hierarchy (\ref{aligncond}).

As we mentioned above, our discussion so far only involved
the scalar masses-squared to leading order
in the SUSY breaking, ${\cal O}(F^2)$.
We do not expect anomaly-mediation to generate
contributions that are higher order in $F$.
To have the correct dimension, such corrections should be of the form 
$\left({F\over \Lambda}\right)^n F^2$,
where $n$ is even and $\Lambda$ is a physical cutoff
associated with the compactification scale.
\footnote{As always in our discussions
in this paper 
our statement is limited to the case that
there are no significant
bulk gravity contributions to the soft masses.  These necessarily contribute
to the masses at the compactification scale and thus alter the UV boundary
conditions on the masses.  As a result the usual formulae for anomaly mediated
soft masses cannot be applied.  In other words, the UV physics does 
not decouple.}

In contrast, when heavy fields of mass $M$ are present in the theory,
with supersymmetry breaking mass splittings,
they generally generate
contributions to the soft masses of the light fields
that are
of order
$F^4/M^2$. 
In the presence of  Yukawa or gauge interactions between heavy and
light fields,
such contributions, which we call ``light-heavy mixing'', 
may even appear at one loop.
Thus, the decoupling of heavy thresholds is not complete
even when the masses of the heavy fields are aligned.
The decoupling holds to leading order in $F^2$,
where there is complete cancellation between
anomaly-mediated contributions and light-heavy mixing
but at order $F^4/M^2$ and higher, there is no
analogous cancellation simply because there is
no anomaly-mediated contribution.

This non-decoupling can again be seen by considering
the regulator fields as above.
We can calculate the contributions of 
both the dynamical fields, whose mass is $M$,
and the regulator fields, whose mass is some physical UV cutoff $\Lambda$
by treating the SUSY breaking parameter $F$ as a small
insertion. However, subleading contributions in $F$
are suppressed by some power of $F/M$ in the case of
the dynamical fields, and by the same power of $F/\Lambda$
in the case of the regulators. For very large $\Lambda$,
the regulator contributions are negligible,
and can no longer cancel the contributions
of the dynamical fields. 
Indeed,  
the $F^4/M^2$ contributions 
will play an
important role in the models we will discuss in the following sections.
We will use these contributions to generate positive masses-squared for
the standard-model sleptons.

\section{A $U(1)$ model}

We will now construct a simple $U(1)$
theory which illustrates the discussion of the
previous section. The $U(1)$ is higgsed,
with some fields becoming massive.
These massive fields then generate one-loop 
${\cal O}(F^4/M^2)$ contributions to the soft masses
of the remaining light fields through gauge and Yukawa
interactions. 
These contributions do not decouple.
Some of the relevant contributions
can be viewed as arising from the D-term.
Different $U(1)$-charged heavy fields receive different anomaly-mediated
contributions to their soft masses, and as a result
a non-zero $U(1)$ D-term is generated. 
Generally, such D-terms do
not decouple from the low energy theory even when the heavy
fields are integrated out~\cite{murayama}. As we will see
in the case of anomaly mediation, the D-term decouples to
leading order in the SUSY breaking $F^2$, but not at higher
orders.

Our $U(1)$ theory consists of the fields $h_\pm$,
$\chi_\pm$, and $l_\pm$, with $U(1)$ charges $\pm 1$, as well as
the gauge singlets $S$, $n_{i=1,2}$,
with the superpotential,
\beq
\label{onestepW}
W =  S\, (\lam_1 \, h_+ h_- - M^2) + 
y_1\, n_1\, h_+ \, \chi_- +
y_2\, n_2\, h_- \, \chi_+ \ ,
\eeq
where $\lam_1$, $y_{i=1,2}$ are couplings.
The first term is needed to break the $U(1)$. As we
will see the last
two terms with different Yukawa couplings, $y_1$ and $y_2$,
are required to generate positive soft masses squared for
some light scalars.
With this superpotential all 
fields except $l_\pm$ acquire mass,
either through the Higgs mechanism or through 
Yukawa interactions. 
We will eventually identify $l_\pm$
with standard model fields.

We assume that the only source of supersymmetry breaking
in the theory is gravitational, through anomaly-mediation.
We can then can keep track of supersymmetry-breaking
effects by rescaling $M \ra M \Phi$, with $\Phi \equiv 1+ F \theta^2$
as before. We also assume that $F \ll M$.
The potential is then given by:
\beq
\label{pot}
V = {\abs {\lambda_1\, h_+ h_- - M^2}}^2 +
\lambda_1^2 \abs{S}^2\, (\abs{h_+}^2 + \abs{h_-}^2)
-2 M^2\, F\, (S+S^*) + \ldots \ ,
\eeq
where we left out terms that involve $n_{1,2}$, $\chi_\pm$,
as we will be interested in a (potentially local) minimum
where these fields do not obtain vevs.
Note that because supersymmetry is broken, there is
a tadpole for the scalar component of $S$, so that it develops
a vev proportional to $F$, 
\beq
\label{svev}
S = -{1\over\lambda_1} F + {\cal O}({F^3\over M^2}) \ ,
\eeq  
whereas $h_\pm$ obtain vevs $h = M/\sqrt{\lambda_1} + {\cal O}(F^2/M)$.
The $U(1)$ is then Higgsed by the $h_\pm$ vevs,
and at tree level we obtain a heavy vector multiplet
with a vector of mass $M_{SUSY}= 2 e h$, a scalar of
mass $m^2 = M_{SUSY}^2 + 2 S^2$, and two fermions
of masses $\sqrt{M_{SUSY}^2 + S^2/4} \pm S/2$, where $e$ is
the $U(1)$ gauge coupling.
In addition, $\chi_-$ mixes with $n_1$ to give a chiral
multiplet with a fermion of mass $y_1 h$ and scalars
of masses-squared $y^2 h^2 \pm y \lambda_1 S h$, and
similarly for the pair $\chi_+$, $n_2$.

At low energies the  theory contains only the fields
$l_\pm$, with no renormalizable interactions.
However, soft masses are generated for the $l_\pm$
scalars through gauge and Yukawa loops containing the
heavy fields.
As discussed in the previous
section, to leading order in supersymmetry breaking,
namely, order $F^2$, these soft masses vanish. We can see
that by working directly in the
low energy theory, which contains no renormalizable interactions,
so that the anomalous dimensions of $l_\pm$ vanish.  
We will return to this point shortly and see
how this can be understood from the point
of view of the full theory.

However, even at one loop, both the heavy gauge multiplet
and the heavy chiral fields generate contributions
to the $l_\pm$ soft masses starting at order 
$F^4/M^2$.\footnote{Recall that unlike $F^2$ terms, such contributions
to the soft masses cannot be read off the wave-function
renormalizations.}
Specifically, the gauge multiplet one-loop contribution
to the $l_\pm$ scalar mass-squared is given by
\beq
\label{mgauge}
 m^2_{\rm gauge} =  q^2\, {1\over 16 \pi^2}\, 4 e^4\, h^2\,
\left[\ln\left(1+ {x^2\over 2 e^2}\right) - 
{8x\over \sqrt{16e^2+x^2}}\, \ln\left(\sqrt{1+{x^2\over 16 e^2}}
+{x\over 4 e}\right)\right] \ , 
\eeq
where $x \equiv {S\over h}$,
and $q$ is the relevant $U(1)$ charge, which for $l_\pm$
is $q=\pm 1$.
This contribution arises purely from the $U(1)$ gauge interactions,

The contribution of the heavy matter fields $\chi_--n_1$ to the $l_\pm$
scalar mass-squared is:
\begin{eqnarray}
\label{my}
m^2_{y} &=& -\, {q\over 64\pi^2}\, y^4 h^2 \,
\left(1+{x^2\over 2e^2}\right)^{-1}
\,\times \cr
&\,&\left[
\left( 2+{x\over y}+{x^3\over y^3}\right)\, 
\ln\left(1+{x\over y}\right) +  
\left( 2-{x\over y}-{x^3\over y^3}\right)\, 
\ln\left(1-{x\over y}\right)
\right] \, ,
\end{eqnarray}
with $y=y_1$, and 
where again, for $l_\pm$,  $q=\pm 1$.
This contribution, although we denote it by
the subscript $y$ for the Yukawa superpotential coupling, arises
from both superpotential interactions
and D-term interactions.
Similarly, the contribution of the heavy matter fields
$\chi_+-n_2$ is
\beq
\label{my2}
m^2_{y_2} = -\, m^2_y\vert_{y=y_2} \ .
\eeq
Note the relative minus sign between~(\ref{my}) and~(\ref{my2}),
which arises from the opposite signs of $\chi_+$ and $\chi_-$.

The contribution of~(\ref{my}) and (\ref{my2}) can
also be understood as arising from the $U(1)$ D-term.
The fields $h_\pm$ obtain soft-masses from
one-loop diagrams with $\chi_\pm-n_{1,2}$
running in the loop.
Since $h_+$ and $h_-$ have different superpotential
couplings ($y_1\neq y_2$), 
their soft masses are also different.
As a result, the $h_+$ and $h_-$ vevs are shifted by
different amounts, so that 
the $U(1)$ D-term is non-zero and
proportional to the difference between the $h_+$ and $h_-$ soft masses
squared.
This non-zero D-term then
leads to soft masses for $l_\pm$, proportional
to their $U(1)$ charges, which are precisely given
by~(\ref{my}). 

Let us now return to the $F^2$ contributions to the
soft masses. As we already mentioned, if we work
directly in the low-energy theory, which contains
only $l_\pm$ with no interactions, the order-$F^2$
anomaly-mediated soft masses, can be read off the (trivial)
wave function renormalizations of $l_\pm$ and therefore vanish.
We could however try to derive this result starting from
the full theory, above the scale of the heavy fields.

In the UV theory, the wave function renormalizations
of $l_\pm$ depend on the gauge coupling, and lead to
non-zero, gauge-coupling dependent soft masses for $l_\pm$,
which we denote by $m_H^2$.
Once we go below the heavy threshold, we have to add
to $m_H^2$ the contribution of the gauge multiplet
to the soft masses which comes from loops involving
the heavy gauge multiplet, $m_G^2$. $m_G^2$ is the analogue
of~(\ref{mgauge}). However, it only comes in at the
two-loop level, since at one loop there can be no
$F^2$ contribution to the soft masses squared~\cite{gr}.
As was shown in~\cite{gr},  $m_G^2$ can also
be read off the supersymmetric wave function renormalizations.
Adding these two contributions, $m_G^2$ precisely cancels $m_H^2$.

The Yukawa-dependent contribution to the $l_\pm$ soft masses
is a bit trickier, since the $l_\pm$ wave function renormalizations
do not depend on the Yukawa couplings.
As discussed above, the Yukawa dependence enters through 
the D-term. As long as $h_+$ and $h_-$ have different
soft masses, the D-term is non-zero, and leads to soft masses
for $l_\pm$. The relevant quantities to consider are then the
$h_\pm$ soft masses. Again, we can read these off the $h_\pm$
wave function renormalizations in the full theory, but
as we integrate out the heavy fields $\chi_+-n_2$ and $\chi_--n_1$,
we obtain $\chi_\pm-n_{1,2}$ loops which precisely cancel 
the original contribution.
Thus, the D-term decouples to order-$F^2$, since the
radiative contributions to the 
$h_\pm$ soft masses vanish in the low-energy theory.
In fact, it would be surprising if  D-terms generated
$F^2$ contributions to soft masses through anomaly-mediation.
We expect to be able to obtain anomaly-mediated $F^2$ soft
masses from wave function renormalizations, which certainly
cannot capture D-term contributions.
It is therefore reassuring to find that the D-term decouples
at order~$F^2$. This is in contrast to the usual case
\cite{murayama} where the leading order SUSY breaking
effects do not decouple. As we have seen, however, there are
non-decoupling effects at order ${\mathcal O} (F^4/M^2)$.

We would eventually like to identify $l_+$ with the standard
model leptons, and to use the contributions we found to its soft mass,
$m^2_{gauge}$ and $m^2_y$, to compensate for the 
negative masses squared which the sleptons obtain in the minimal
anomaly-mediated scenario~\cite{rs}. 
However, $m^2_{gauge}$ and $m^2_y$ are proportional
to $F^4/M^2$, so that would involve tuning the ratio of two
unrelated scales, $F$ and $M$. In the next section
we will present a model in which a ``large'' scale 
$M \sim F/\lam_0$, where $\lam_0$ is some Yukawa coupling,
is generated dynamically.

\section{A two step model}

Our goal is to find a mechanism which would naturally
generate a mass scale which is somewhat larger than the SUSY
breaking scale, yet parametrically is of the same order. In
fact the model considered in the previous section suggests a
way to achieve this goal.
Our main observation is that the field $S$ 
obtained a small vev of the order
$F/\lam_0$. For relatively small $\lam_0$ this scale may be
sufficiently large, so that the $S$ threshold is
supersymmetric, yet it is not parametrically large, and can
therefore play the role of the mass scale $M$ of the
previous section. Thus we are lead to the modification of
our model which has the following 
superpotential\footnote{This superpotential is the
most general one preserving a $U(1)_R$ symmetry.}
\beq
\label{twostepW}
W = X (\lam_0 n^2 - \tilde M^2) + S\, (\lam_1 h_+ h_- - \lam_2 X^2) + 
y_1 n_1\, h_+ \, \chi_- +
y_2 n_2\, h_- \, \chi_+ \ .
\eeq
We now describe the role of the various
terms in this superpotential.
The last three terms are exactly the same as
in the superpotential (\ref{onestepW}) with the 
substitution
$M^2 \ra \lam_2 X^2$. This substitution is justified since
$X$ is heavy. Moreover, all other fields in the above
superpotential become heavy at the minimum at least in the
sense of eq.~(\ref{aligncond}).
If $X$ obtains a non-vanishing vev, this sector
of the theory will generate 
soft masses for $l_\pm$ 
exactly as in the previous section.

The first term in the superpotential~(\ref{twostepW})
is designed to produce an  $X$ vev of the order $F/\lam_0$. 
The large scale $\tilde M$ is necessary to generate this vev 
as well as other field vevs  and eventually leads 
to the breaking of the $U(1)$ gauge
symmetry. Such a large mass parameter may
be naturally generated by various strong coupling effects in
the microscopic theory. It is important, however, that our
results are insensitive to the precise value of $\tilde
M$.

Let us now analyze the model. Upon supersymmetry
breaking the scalar potential has the form (neglecting the
couplings to $\chi_\pm$)
\beq
\label{twostepV}
\begin{array}{c}
V = \abs{ \lam_1 h_+ h_- - \lam_2 X^2}^2 + \abs{ \lam_1 S
h_+}^2 + \abs{ \lam_1 S h_-}^2 + \\
\abs{ - 2 \lam_2 S X + \lam_0 n^2 - \tilde M^2}^2 + \abs{2 \lam_0 X n}^2 - 
(2 \tilde M^2 F X + c.~c.) ~.
\end{array}
\eeq

We wish to separate our analysis in two stages. First we set
$\lam_2=0$ and find vevs for $n$ and $X$ using our results
from the previous section. Since these fields
are heavy at the minimum we then assume
that their vevs are not significantly shifted in the full
theory. 
We will discuss under which conditions this
assumption is true shortly. 
Then we turn on $\lam_2$ and 
integrate out the heavy fields $X$, and $n$, and
then we consider an effective theory for the lighter fields
$S$, $h_+$ and $h_-$. Using 
our results from the previous section  we find
that $X = F/\lam_0$. It is also easy to find that
$F_X = F^2/\lam_0$.
The superpotential for the lighter fields now acquires the form
(\ref{onestepW}) with the substitution 
$M= \sqrt{\lam_2} X$.
Note that
the value of mass scale $\tilde M$ is indeed irrelevant for the
dynamics of the lighter degrees of freedom, $S$ and $h_\pm$. 
We now have to check that the mass parameter
$M$ can be promoted to a superspace valued background
superfield with an F-term expectation value $M F$. Indeed,
\beq
\label{Xalign}
\sqrt{\lam_2} X = \sqrt{\lam_2}\frac{F}{\lam_0} + 
\sqrt{\lam_2} \frac{F^2}{\lam_0} \theta^2 =
\frac{\sqrt{\lam_2} F}{\lam_0} (1 + F \theta^2)   ~.
\eeq
Thus we can use our results from the previous section
for the soft masses of $l_\pm$ by substituting
$M\ra \sqrt{\lam_2} F/\lam_0$.

Finally, let us turn to the conditions for the existence of
the desired minimum in the model~(\ref{twostepW}). 
We note
that the model possesses a flat direction
parameterized by the $S$ vev. In the
limit of unbroken SUSY and in the region of the moduli space
of interest $S$ is heavy and can not acquire a vev until
$h_+\sim h_- \ra 0$. However, when we turn SUSY breaking on,
it is possible that one mass eigenvalue for the two
real fields in the $S$ supermultiplet will become negative, and
the local minimum will not exist. We should, therefore,
require that the F-type SUSY violating masses for $S$ are
much smaller that the supersymmetric contribution to the
mass. For the supersymmetric mass we have $m^2_S= 2
\lam_0^{-2} \lam_1 \lam_2 F^2$, while the soft mass is $m_F^2 =
\sqrt{2}\lam_1 F_h = \sqrt{2 \lam_0^{-2} \lam_1\lam2}
F^2$.
We easily see that a local minimum exists if 
\beq
\label{ratio}
\frac{\lam_0}{\sqrt{\lam_1 \lam_2}} \ll 1 \ .
\eeq
Note that the combination of couplings in (\ref{ratio}) is
exactly the quantity $x$ which enters formulae
(\ref{mgauge}) and (\ref{my}) for the soft masses.
We performed a numerical minimization of the scalar potential,
and verified that the local minimum exists for a range of
parameters when the ratio in~(\ref{ratio}) is of order
or smaller than $0.1$.

\section{Correcting the slepton masses}

As was pointed out in \cite{rs}, the minimal AM scenario 
in which the sole origin of the superpartner masses 
is anomaly mediation
gives negative slepton masses squared. Thus extra contributions 
to the slepton masses squared are required.
We will now use the model we constructed in the
last two sections to generate positive contributions to these masses.
To do that we augment the SM gauge group by the $U(1)$ of the model
we described earlier, and charge the lepton fields under this $U(1)$,
with $U(1)$ charge $+1$. Our starting point is then the theory
described in Section~3, with six copies of the field $l_+$ corresponding to
the six SM lepton fields: $l_+^{i} =(1,2,-1/2,+1)$,
$l_+^{i+3} =(1,1,-1,+1)$, where $i=1\ldots 3$ is a generation index,
and the parenthesis indicate
the SM$\times U(1)$ representation. Similarly, we also take six copies
of the fields $\chi_-$, with $\chi_-^i =(1,2,1/2,-1)$,
and $\chi_-^{i+3} =(1,1,1,-1)$, and six copies of the field $n_1$
with $n_1^i =(1,2,-1/2,0)$, and $n_2^{i+3}=(1,1,1,0)$.
It is easy to check that with this field content there are no
SM$\times U(1)$ anomalies.
The field content of the model is summarized in Table~1.
%%%%%%%%%%%%%%%%%%%%%%%%%%%%%%%%%%%%%%%
%  TABLE 1                            %
%%%%%%%%%%%%%%%%%%%%%%%%%%%%%%%%%%%%%%%
%
\begin{table} \begin{center}
{\centerline{Table 1: The field content of the model}}
{\centerline{(we do not show SM fields that are neutral under the $U(1)$).}}
\label{table1}
\begin{tabular}{c|c|c} \hline \hline
$\ $ & SM & $U(1) $ \\
\hline
$S$ & $(1,1,0)$ & 0 \\
$X$ & $(1,1,0)$ & 0 \\
$n$ & $(1,1,0)$ & 0 \\
$h_+$ &$(1,1,0)$ & 1 \\
$h_-$ &$(1,1,0)$ & -1 \\
\hline
$L\equiv l_+^{i=1..3}$ & $(1,2,-1/2)$& 1  \\
$ \chi_-^{i=1..3}$ & $(1,2,1/2)$& $-1$  \\
$n_1^{i=1..3}$ & $(1,2,1/2)$& 0  \\
${\bar l}\equiv l_+^{i=4..6}$ & $(1,1,-1)$& 1  \\
$ \chi_-^{i=4..6}$ & $(1,1,,1)$& $-1$  \\
$n_1^{i=4..6}$ & $(1,1,-1)$& 0  \\
\hline
\end{tabular}
\end{center}
\end{table}
%%%%%%%%%%%%%%%%%%%%%%%%%%%%%%%%%%%%%%%
%
The superpotential is then given as in~(\ref{twostepW}),
\beq
\label{twostepWr}
W = X (\lam_0 n^2 - \tilde M^2) + S\, (\lam_1 h_+ h_- - \lam_2 X^2) + 
y_1 \sum_{i=1}^6 n_1^i\, h_+ \, \chi_-^i \ ,
\eeq
except that so far we have not added the fields $l_-$, $\chi_+$
and $n_2$, so that the superpotential term containing them
does not appear.
The fields $\chi_-^i$, $n_1^i$  become
heavy, with mass $y_1 h$. 
As discussed in Section~2, the sleptons then obtain the following soft masses
\beq
\label{mslepton}
m_{slepton}^2 = m^2_{gauge} + 9 m_y^2\vert_{y=y_1} \ ,
\eeq
where $m^2_{gauge}$ and $m^2_y$ are given in~(\ref{mgauge})
and~(\ref{my}) with $q=1$.

Let us now discuss the different mass contributions.
Recall that the heavy matter contribution to the $l_+$ mass, 
$m^2_y$, is proportional to the charge of $l_+$,
whereas  $m^2_{gauge}$ is proportional to the square
of the charge. In addition, $m^2_{gauge}$ is always
negative. Thus, if we want to generate positive masses squared for
all sleptons, they all have to have $U(1)$ charges of the same
sign. Thus, we choose all leptons to have $U(1)$ charge $+1$,
and identify them with the fields $l_+$. To cancel anomalies,
we then add fields of the type $\chi_-$, of 
$U(1)$ charge $-1$, and $n_1$. These fields then become heavy,
and at one loop generate the contribution $ m_y^2\vert_{y=y_1}$.
Unfortunately, however, this contribution is also negative!
Examining eq.~(\ref{my2}), we see that in order to get
a positive contribution, we need heavy fields of opposite
$U(1)$ charge running in the loop, that is, fields of
the type $\chi_+$ (and their $n_2$ partners). 
Again anomaly considerations then require the presence of
additional fields $l_-$. Unlike the $\chi$'s and the $n$'s,
these fields remain light, so that the simplest possibility is to 
identify them with some of the SM fields. We are therefore led
to charging additional SM fields under the $U(1)$, 
with charges that are opposite in sign to the lepton charges.
We will now discuss two possibilities of doing so.
One in which the down antiquarks have charge $-1$, and the
other in which the first and second generation quarks
have charge $-1$. While we will be able to generate positive
masses squared for the sleptons, we see from the required
matter content that 
probably the most disappointing aspect of our model is
that it can not be made consistent with grand unification. 
As explained above,
the lepton fields all have the same $U(1)$ charge, and
some other SM fields should have the opposite $U(1)$ charge.
This automatically excludes both $SU(5)$ and $SO(10)$ unification.
Moreover, the additional matter fields do not come in GUT
representations, and even gauge coupling unification
requires the 
introduction of extra matter at intermediate scales.

\subsection{$U(1)$ charged leptons and down antiquarks}

As discussed above, in order to obtain positive slepton
masses-squared, we need heavy fields of positive
$U(1)$ charge running in the loop.
We can achieve that by identifying the SM down antiquarks
with the fields $l_-$.
Thus, there are three copies of the field $l_-$:
 $l_-^i=({\bar 3}, 1,-1/3,-1)$, which are accompanied
by three copies of the field $\chi_+$ with 
$\chi_+^i=(3, 1,1/3,1)$ and three copies of $n_2$
with $n_2^i=({\bar 3}, 1,-1/3,0)$.
This additional field content  is summarized in Table~2.
%%%%%%%%%%%%%%%%%%%%%%%%%%%%%%%%%%%%%%%
%  TABLE 2                            %
%%%%%%%%%%%%%%%%%%%%%%%%%%%%%%%%%%%%%%%
%
\begin{table} \begin{center}
{\centerline{Table 2: Additional fields: down antiquarks}}
{\centerline{(we do not show SM fields that are neutral under the $U(1)$).}}
\label{table2}
\begin{tabular}{c|c|c} \hline \hline 
$\ $ & SM & $U(1) $ \\
\hline
$\ $ & $\ $ & $\ $\\
${\bar d}\equiv l_-^{i=1..3}$ & $({\bar 3},1,-1/3)$ & $-1$\\
$\chi_+^{i=1..3}$ & $(3,1,1/3)$ & $-1$\\
$n_2^{i=1..3}$ & $({\bar 3},1,-1/3)$ & 0\\
\hline
\end{tabular}
\end{center}
\end{table}
%%%%%%%%%%%%%%%%%%%%%%%%%%%%%%%%%%%%%%%
%

The superpotential is then given exactly as in~(\ref{twostepW}).
For convenience we rewrite it here:
\beq
\label{twostepWrr}
W = X (\lam_0 n^2 - \tilde M^2) + S\, (\lam_1 h_+ h_- - \lam_2 X^2) + 
y_1 \sum_{i=1}^6 n_1^i\, h_+ \, \chi_-^i +
y_2 \sum_{i=1}^3  n_2^i\, h_- \, \chi_+^i \ .
\eeq
The fields $\chi_-^i$, $n_1^i$ ($\chi_+^i$, $n_2^i$) all become
heavy, with mass $y_1 h$ ($y_2 h$). The sleptons and
down antiquarks obtain the following soft masses
\begin{eqnarray}
\label{msleptonmd}
m_{slepton}^2 &=& m^2_{gauge} + 9 m_y^2\vert_{y=y_1} 
- 9 m_y^2\vert_{y=y_2} \ ,\cr
m_d^2 &=& m^2_{gauge} - 9 m_y^2\vert_{y=y_1} + 9 m_y^2\vert_{y=y_2} \ .
\end{eqnarray}
Note that because the sleptons and down quarks have 
opposite $U(1)$ charges, their soft masses obtain opposite
contributions from the heavy matter fields.

As explained in the beginning of this section, $m^2_{gauge}$
and $m^2_y$ are negative. Thus, to get a positive contribution
to the slepton masses squared, the third term in the first
line in~(\ref{msleptonmd}) should overcome the first two.
We also point out that
$m_{gauge}^2$, $m_{y_1}^2$, and $m_{y_2}^2$ all start at
order $h^2 x^4 = s^4/ h^2$. At this order, the dependence on
the couplings $e$, $y_i$ drops out, so that the ${\cal O} (x^4)$
terms cancel between $m_{y_1}^2$ and $m_{y_2}^2$.

It is  clear from~(\ref{msleptonmd}) that as long as the sleptons
obtain a positive contribution to the mass squared, the
down-type squarks get a negative contribution. We then have to ensure
that the new contribution to the slepton mass is bigger than the
one generated directly by anomaly-mediation, whereas the new contribution
to the down squark mass is smaller than the one generated by anomaly-mediation.
Thus, we need,
\begin{eqnarray}
\label{mcondition}
m_{slepton}^2 > \abs{m_{slepton, AM}^2} \ ,
~~~~~\abs{m_{d}^2} < m_{d, AM}^2 \ ,
\end{eqnarray}
where $m_{slepton, AM}^2$ and $m_{d, AM}^2$ are the slepton 
and down squark masses generated by anomaly-mediation
in the absence of any heavy thresholds,
\beq
\label{pureAMld}
m_{slepton, AM}^2 \sim 10^{-3}\, {F^2\over 16\pi^2}  \ , ~~~~~~~
m_{d, AM}^2 \sim 10^{-1}\, {F^2\over 16\pi^2}  \ .
\eeq
Thus, the different couplings need to be tuned
to satisfy this relation. It turns out that the tuning
required is not drastic. First, no large hierarchy is required
between the couplings $e$, $y_1$ and $y_2$ to obtain a
positive $m_{slepton}^2$. Second, to satisfy~(\ref{mcondition}),
the coupling $\lam_0$, which controls the size of $h$,
can vary within an overall factor of around five.
 
We now turn to consider the SM Yukawa couplings. Since
both the $SU(2)$-doublet and -singlet leptons have $U(1)$
charge $+1$, and the down quarks have $U(1)$ charge $-1$,
the lepton and down Yukawas are not neutral under the
$U(1)$. These Yukawas can however arise from non-renormalizable
terms,
\beq
\label{yukawas}
{1\over M}\, h_+\, H_d\, Q \,{\bar d} 
+ {1\over M^2}\, h_-^2\, H_d\, L\, {\bar  l} \ ,
\eeq
where $M$ is some higher scale.
We are thus led to a model with nontrivial flavor structure,
where the up Yukawas have no suppression, the down
Yukawas are suppressed by one power of $h/M$, and the
lepton Yukawas are suppressed by two powers of $h/M$.

\subsection{$U(1)$-charged leptons and first generations quarks}

An alternative to assigning $U(1)$ charge $-1$ to the
SM down-type antiquarks, is to assign charge $-1$ to the
first and second generation doublet quarks.
We then have fields $l_+^i$, $\chi_-^i$ and $n^i$ as in the
previous subsection, as well as the the first and
second generation quarks, which we identify with
$l_-^{i=1,2} =(3,2,1/6,-1)$,  $\chi_+^{i=1,2}=(3,2,-1/6,1)$
and $n_2^{i=1,2}=(3,2,1/6,0)$.
We summarize the additional field content of this model in Table~3.
%%%%%%%%%%%%%%%%%%%%%%%%%%%%%%%%%%%%%%%
%  TABLE 3                            %
%%%%%%%%%%%%%%%%%%%%%%%%%%%%%%%%%%%%%%%
%
\begin{table} \begin{center}
{\centerline{Table 3: Additional fields: 1$^{\rm st}$ and
2$^{\rm nd}$ generation quarks}}
{\centerline{(we do not show SM fields that 
are neutral under the $U(1)$).}}
\label{table3}
\begin{tabular}{c|c|c} \hline \hline
$\ $ & SM & $U(1) $ \\
\hline
$Q\equiv l_-^{i=1,2}$ & $(3,2,1/6)$ & $-1$\\
$\chi_+^{i=1,2}$ & $({\bar 3},2,-1/6)$ & 1\\
$n_2^{i=1,2}$ & $(3,2,1/6)$ & 0\\
\hline
\end{tabular}
\end{center}
\end{table}
%%%%%%%%%%%%%%%%%%%%%%%%%%%%%%%%%%%%%%%
%

The sleptons and first and second generation squarks now
obtain the following  contributions to their masses-squared:
\begin{eqnarray}
\label{msleptontwo}
m_{slepton}^2 &=& m^2_{gauge} + 9 m_y^2\vert_{y=y_1} 
- 12 m_{y_2}^2\vert_{y=y_2} \ ,\cr
m_d^2 &=& m^2_{gauge} - 9 m_y^2\vert_{y=y_1} 
+ 12 m_y^2\vert_{y=y_2} \ .
\end{eqnarray}
Again, the third term on the first line gives a positive contribution, 
with the two other
contributions negative.
As in the last subsection, we can tune the various couplings
so that the total slepton mass is positive, and
the squark mass receives only a small (negative) correction.

The lepton Yukawa couplings again arise from non-renormalizable
operators, and are suppressed by two powers of $h/M$.
As for the quark Yukawa couplings, the third generation term
is not suppressed, whereas the first two generations are
suppressed by one power of $h/M$.

Note that, unlike in the previous subsection, the soft masses
are no longer flavor-blind: the soft masses of the first two 
generation squarks receive negative corrections and are smaller than 
the soft masses of the third generation squarks. However, these
corrections can be chosen to be small, with no severe tuning of 
parameters.
In addition, constraints on FCNC processes which involve the third 
generation
are typically weaker.

\section{The $\mu$-term from anomaly mediated SUSY breaking}

We now turn to the $\mu$-term problem in AMSB. As has
been noted in~\cite{rs} this problem is much less severe
than in gauge mediated models, and indeed several mechanisms
generating $\mu$ and $B$ terms of the correct order of
magnitude have been proposed recently
\cite{rs,rp,luty}. Here we propose a solution to this
problem based on the use of the  higher order (in SUSY
breaking) correction as well as the observation that in AMSB
models it is easy to generate a scale which is somewhat
larger than gravitino mass, $m_{3/2} \sim F$.

We introduce the superpotential
\beq
\label{muW}
W = \lambda_H S H_u H_d + \lambda_S S^3 + \lambda_N S N^2 +
M N^2 \ ,
\eeq
where the mass parameter $M$ is assumed to be generated
dynamically as in Section 4, $M = F/y$ for some coupling 
constant $y$. In addition, the scalar components of $N$ have
soft tree level contributions to their masses, $MF$.
Generally, the Higgs boson vevs lead to a vev (and therefore
an effective $\mu$ term) for $S$ through the superpotential
(\ref{muW}). We will argue shortly that such a contribution
does not affect the
conclusions we will draw, and therefore, will neglect it
throughout our discussion.
 
Anomaly mediation generates a positive
contribution to the $S$ singlet mass
squared of the order
\beq
m_{AM}^2 \sim \frac{\lambda_S^4 + \lam_H^4}{(16 \pi^2)^2}\, F^2 \ .
\eeq
Here and throughout this section we omit some order one
numerical coefficients.
On the other hand, a one loop negative mass squared for the
singlet is generated due to the non-decoupling of the heavy
states,
\beq
\label{mnegative}
m_{F^4}^2 \sim - \frac{\lambda_N^2}{16 \pi^2}\, \frac{F^4}{M^2} =
- \frac{\lambda_N^2}{16 \pi^2}\, y^2\,  F^2 \ .
\eeq
It is easy to see that the singlet mass will be negative as
long as\footnote{Remember that $N$ is heavy, and as a result
$\lam_N$ does not contribute to the $S$ mass at order $F^2$.}
\beq
\label{condition}
\lam_N^2 y^2 > \frac{\lam_H^4 + \lam_S^4}{16 \pi^2} \ ,
\eeq
where the right hand side indicates the order of magnitude only.

As a result both the scalar and the auxiliary components of $S$
acquire vevs 
\begin{eqnarray}
S &\sim& \frac{1}{4 \pi}\frac{\lambda_N y}{\lam_S}\,  F \ ,\\ \nonumber
F_S &\sim& \frac{1}{16 \pi^2}
\frac{\lambda_N^2 y^2 }{\lam_S}\, F^2 \ .
\end{eqnarray}

Substituting these vevs into the superpotential (\ref{muW})
we find that both $\mu$ and $B$ are generated
\begin{eqnarray}
\label{muB}
\mu &\sim& \frac{1}{4 \pi}
\frac{\lam_H}{\lam_S} y \lambda_N F \ ,\\ \nonumber
B &\sim& \frac{1}{16 \pi^2}\frac{\lambda_H}{\lam_S}
 y^2\lambda_N^2 F^2 \ .
\end{eqnarray}
After $S$ acquires a vev there is an additional contribution to 
$B$ arising from the $S\,H_u\,H_d$ A-term, however, it is negligible
when~(\ref{condition}) is satisfied.
It is easy to see that $B$ and $\mu^2$ are of the same order
if 
\beq
\label{sameorder}
\lam_H/\lam_S \sim {\mathcal O}(1) \ .
\eeq
We further need to require that the $\mu$ term is of the order
of the weak scale,
\beq
 \frac{1}{4 \pi}
\frac{\lam_H}{\lam_S} y\lambda_N F \sim 
\frac{\alpha_2}{4 \pi} F \ .
\eeq
This requirement together with (\ref{sameorder}) gives a
condition on two Yukawa couplings
$\lam_N y \sim \alpha_2$.
We note that this condition is quite compatible with
the requirement that $S$ has a negative mass squared.

Finally we observe that the Higgs vevs generate a mass term for
$S$. This mass contribution is below 
the negative mass~(\ref{mnegative})
by roughly $\lam_H \, H_U/\mu$. Since $\lam_H$ can be arbitrary as
long as it is comparable with $\lam_S$, such a contribution
is small compared to the negative mass generated by
non-decoupling effects (which we can arrange to be between
electroweak scale and $1 {\mathrm TeV}$). Even in the case
$\lam_H \sim 1$, our qualitative conclusions remain valid,
and both a  $\mu$ and a $B$ term of the correct order of magnitude
are generated.
 
Having established that eq.~(\ref{muB}) gives a leading
contribution to $\mu$ and $B$ it is possible to show that
the physical phase $\phi = {\rm arg} (B \mu^*
M_\lambda^*)$ vanishes. Here $M_\lambda$ is the gaugino
mass. Thus, this sector of the theory does
not lead to a SUSY CP problem.

To conclude our discussion of the $\mu$ term, we observe that the
superpotential (\ref{muW}) could be introduced in a gauge
mediated model, with $N$ being a messenger field. However, in
calculable models of gauge mediation the scale of
supersymmetry breaking is relatively large, while the messenger
mass is at most suppressed by several loop factors relative
to this scale. As a result the higher order contributions
used here are too small to generate electroweak
scale parameters.~\footnote{This is not a problem in
strongly coupled gauge mediated models with a low  
SUSY breaking scale. However, such models are non-calculable, and
it is not possible to quantitatively analyze their
spectrum.}
In principle it is possible to generate a
small mass for the messengers, however, this requires the
introduction of a quite complicated structure and explicit 
mass scales, 
unlike with anomaly mediation where a mass scale somewhat
larger than the SUSY breaking scale of the visible sector can
naturally be generated.

\section{Conclusions}

In this paper, we studied the decoupling of heavy
thresholds in theories with anomaly-mediated supersymmetry
breaking. To leading order in the supersymmetry breaking,
such thresholds decouple. That is, the anomaly-mediated
supersymmetry breaking terms at some low scale are
independent of whether or not there are supersymmetric 
thresholds above that scale. These soft terms are thus
quite robust.

It is possible to see this decoupling in several ways.
For example, it can be understood as a cancellation
between the following two quantities: The first is the contribution of 
the heavy fields to the anomaly mediated soft terms
in the full theory. Recall that the AM soft terms depend
on the beta function of the theory, which in the
full theory reflects the presence of the heavy fields.
The second is the direct radiative contribution, through
gauge or Yukawa interactions, of the heavy fields
to the soft terms of the light fields. This
contribution is generated when the heavy fields are
integrated out. These two contributions cancel exactly,
so that below the scale of the heavy fields, they leave
no trace on the soft terms at leading order in SUSY breaking.
That these two contributions exactly cancel can be seen
on a case by case basis, but it is most simply seen from the
fact that the direct gauge- or Yukawa-mediated contributions
of the heavy fields can be read off the wave function 
renormalizations~\cite{gr,gr1} in precisely the same
way as the AM contributions.

Alternatively, the decoupling can be seen as a cancellation
between ``real'' fields and their regulators.

When do heavy threshold not decouple? One
obvious possibility is that they are not truly
heavy~\cite{rp}. That is, there is some light modulus associated
with the heavy threshold whose mass comes mainly from SUSY 
breaking effects.
It is worth pointing out that approximate decoupling persists even 
for a modulus much lighter than the SUSY breaking scale $F$ so long as 
its mass is primarily determined by supersymmetric parameters.

However, there is additional non-decoupling even when
all the heavy fields are truly heavy. That is because
the purely anomaly-mediated soft terms only appear
at leading order in the SUSY breaking. For example,
scalar masses squared are order $F^2$. In contrast,
as we integrate out some heavy fields, they give
direct contributions, again through loops, to the soft
terms to all orders in the SUSY breaking. Scalar
masses squared now have contributions of order $F^4/M^2$,
where $M$ is the heavy threshold.
Moreover, these contributions  typically appear at lower
order in the loop expansion. For scalar masses squared,
they can appear at one-loop, whereas the AM soft masses
are two-loop contributions.

Having established that heavy supersymmetric thresholds
do affect the soft terms at order $F^4$, we then use
this fact to generate positive slepton masses.
If the only source of SUSY breaking in the SM is anomaly
mediation, and if there are no supersymmetric thresholds,
the slepton masses squared are negative. 
However, we can charge the leptons (and another subset
of SM fields) under a new $U(1)$ gauge symmetry, which
is broken at a scale somewhat above the visible sector
SUSY-breaking scale, and add some fields 
that obtain supersymmetric masses. In the presence
of anomaly-mediation, these heavy fields also acquire
SUSY-breaking masses, and contribute to slepton 
masses at order $F^4$.
Interestingly, we are led to a model with some non-trivial
flavor structure.
Unfortunately, this model is not consistent with grand
unification.

As another model building application, we used the
$F^4$ contributions of heavy supersymmetric thresholds
to generate acceptable $\mu$- and $B$-terms.
As we saw, this can be done quite simply in models of
anomaly mediation, unlike in the case of gauge-mediation.
In the latter case, an acceptable $\mu$ term typically leads
to a $B$ term that is too large.

In both these model building examples, we also use
a simple mechanism that dynamically generates a scale
that is naturally somewhat above the SUSY breaking scale
through anomaly-mediation.
We expect this fact to be useful for further model
building applications.

\bigskip\bigskip

We would like to thank Lisa Randall for many
useful discussions, and for collaboration during
the  early stages of this work. We also thank
Michael Dine, Jonathan Feng, and Yossi Nir for discussions.
This work was supported in part by NSF grant
\#PHY-9802484, and DOE grants \#DF-FC02-94ER40818 
and \#DE-FC02-91ER40671.

\nc{\ib}[3]{ {\em ibid. }{\bf #1} (19#2) #3}
\nc{\np}[3]{ {\em Nucl.\ Phys. }{\bf #1} (19#2) #3}
\nc{\pl}[3]{ {\em Phys.\ Lett. }{\bf #1} (19#2) #3}
\nc{\pr}[3]{ {\em Phys.\ Rev. }{\bf #1} (19#2) #3}
\nc{\prep}[3]{ {\em Phys.\ Rep. }{\bf #1} (19#2) #3}
\nc{\prl}[3]{ {\em Phys.\ Rev.\ Lett. }{\bf #1} (19#2) #3}

\end{document}